\documentstyle[graphics]{mn}
\newcommand{\etal}{{\it et al.}~}

\newcommand{\bc}{\begin{center}}
\newcommand{\be}{\begin{equation}}
\newcommand{\ee}{\end{equation}}
\newcommand{\ec}{\end{center}}

\newcommand{\Mpc}{\rm Mpc}

\newcommand{\tree}{TREESPH }
\newcommand{\lya}{Lyman-$\alpha$~}
\newcommand{\eg}{{\it e.g.~}}
\newcommand{\ie}{{\it i.e.~}}
\renewcommand{\H}{{\mbox{${\rm HI~}$}}}

\newcommand{\ltsima}{\mbox{$\; \buildrel < \over \sim \;$}}
\def \simlt{\lower.5ex\hbox{\ltsima}}            
\def \gtsima{\mbox{$\; \buildrel > \over \sim \;$}}
\def \simgt{\lower.5ex\hbox{\gtsima}}            

\title[Low-redshift evolution of the \lya Forest]
{The low-redshift evolution of the \lya Forest}

\author[T. Theuns \etal]
{Tom Theuns$^{\,1,2}$, A. P. B. Leonard$^{\,2}$ and George Efstathiou$^{\,1,2}$\\
$^{\,1}$ Institute of Astronomy, Madingley Road, Cambridge CB3 0HA, UK\\
$^{\,2}$ Department of Physics, Astrophysics, University of Oxford,
Keble Road, Oxford OX1 3RH, UK
}

\begin{document}

\maketitle

\begin{abstract}
The low-redshift evolution of the intergalactic medium is investigated
using hydrodynamic cosmological simulations. The assumed cosmological
model is a critical density cold dark matter universe. The imposed
uniform background of ionizing radiation has the amplitude, shape and
redshift evolution as computed from the observed quasar luminosity
function by Haardt \& Madau. We have analysed simulated \lya spectra
using Voigt-profile fitting, mimicking the procedure with which quasar
spectra are analysed. Our simulations reproduce the observed evolution
of the number of \lya absorption lines over the whole observed interval
of $z=0.5$ to $z=4$. In particular, our simulations show that the
decrease in the rate of evolution of \lya absorption lines at $z\le 2$,
as observed by {\em Hubble Space Telescope}, can be explained by the
steep decline in the photo-ionizing background resulting from the rapid
decline in quasar numbers at low redshift.
\end{abstract}


\section{Introduction}
Neutral hydrogen in the intergalactic medium produces a forest of \lya
absorption lines blueward of the \lya emission line in quasar spectra
(Lynds 1971). Observations of quasars at redshifts $z>2$ show that
there is strong cosmological evolution in the number of \lya lines,
which can be characterised by a power law $dN/dz \propto (1+z)^\gamma$
(Sargent \etal 1980), where $N$ is the number of lines above a
threshold rest-frame equivalent width $W$ (typically
$W>0.32\AA$). Studies at high resolution using the Keck telescope find
$\gamma=2.78\pm 0.71$ for $2<z<3.5$ (\eg Kim \etal 1997). At even
higher $z$ the evolution is still stronger: Williger \etal (1994) find
$\gamma>4$ for $z>4$ using the CTIO 4m telescope. In contrast at low
redshifts, observations using the {\em Hubble Space Telescope} find
much less evolution, $\gamma=0.48\pm0.62$ for $z<1$ (Morris \etal 1991,
Bahcall \etal 1991, 1993, Impey \etal 1996).

Recently, hydrodynamic simulations of hierarchical structure formation
in a cold dark matter (CDM) dominated universe have been shown to be
remarkably successful in reproducing this \lya forest in the redshift
range $z=4\rightarrow 2$ (Cen \etal 1994, Zhang, Anninos \& Norman
1995, Miralda-Escud\'e \etal 1996, Hernquist \etal 1996, Wadsley \&
Bond 1996, Zhang \etal 1997, Theuns \etal 1998). These simulations show
that the weaker \lya lines (neutral hydrogen column density $N_\H\le
10^{14}$ cm$^{-2}$) are predominantly produced in the filamentary and
sheet-like structures that form naturally in this structure formation
scenario. Velocity structure in these lines is often due to residual
Hubble flow since many of the absorbing structures are expanding. In
contrast, the stronger lines ($N_\H\ge 10^{16}$ cm$^{-2}$) tend to
occur when the line of sight passes near a dense virialised halo.

Over the redshift range investigated in these simulations a
photo-ionizing background close to that inferred from quasars (Haardt
\& Madau 1996) is required to explain the properties of the \lya
forest. In fact, although most simulations have assumed a critical
density, scale-invariant CDM universe, other variants of the CDM model
provide acceptable fits with relatively small changes to the ionizing
background (Cen \etal 1994, Miralda-Escud\'e \etal 1996). The general
success of CDM-like models in explaining the high redshift ($z\ge 2$)
properties of the \lya forest is impressive.

In this {\em Letter} we investigate using hydrodynamic simulations
whether a CDM universe with a photo-ionizing background dominated by
quasars can explain the observed transition in the cosmological
evolution of the number of \lya lines at $z\le 2$.

\section{Simulation}
We model the evolution of a periodic, cubical region of a critical
density Einstein-de Sitter universe ($\Omega=1$,
$\Omega_\Lambda=0$). We use a simulation code based on a hierarchical
P3M implementation (Couchman 1991) for gravity and smoothed particle
hydrodynamics (Lucy 1977, Gingold \& Monaghan 1977, see \eg Monaghan
1992 for a review) for hydrodynamics (Theuns \etal 1998). The comoving
size of the simulation box is $L/(2h)$ \Mpc, where the Hubble constant
today is written as $H_0=100 h$ km s$^{-1}$ Mpc$^{-1}$. We will assume
$h=0.5$ throughout and describe simulations with $L=5.5$ and
$22.22$\Mpc. A fraction $\Omega_Bh^2=0.0125$ of the matter density is
assumed to be baryonic, consistent with limits from nucleo-synthesis
(Walker \etal 1991, but note the continuing debate on the deuterium
abundance derived from quasar spectra which might favour higher values
$\Omega_B h^2=0.019$, see \eg Burles \& Tytler 1997 and references
therein). The rest of the matter is assumed to be in the form of cold
dark matter. To generate initial conditions for the simulations at the
starting redshift $z=50$ we use the fit to the cold dark matter linear
transfer function from Bardeen \etal (1986) and normalise it such that
the linearly extrapolated value of $\sigma_{8}=0.7$ at the present
day. These simulations use $64^3$ particles of either species hence the
gas mass resolutions are $1.5\times 10^8 M_\odot$ and $2.2\times 10^6
M_\odot$ for $L=22.22$ and $L=5.5$Mpc, respectively.

Gas in these simulations is ionized and photo-heated by an imposed
uniform background of ionizing photons assumed to originate from
quasars as computed by Haardt \& Madau (1996). This flux is redshift
dependent, mimicking the evolution of the quasar luminosity
function. Gas can also cool by interacting with microwave background
photons and through collisional cooling. The detailed rates for all
these processes as a function of temperature are taken from Cen (1992)
with some minor modifications. We assume ionization equilibrium
throughout and use a helium abundance of $Y=0.24$ by mass. We have
compared in detail the results from our code against the published
results of \tree (Hernquist \etal 1996, Croft \etal 1997) and find
excellent agreement (see Theuns \etal 1998 for more details of these
comparisons and for a description of our code and cooling rates).

The effective mean optical depth $\bar\tau$ from the simulations, with
this set of parameters, is significantly lower than the observed
value. Consequently, we have reduced the amplitude $\Gamma_\H$ of the
ionizing radiation given by Haardt \& Madau (1996) by a factor of
two. Here, $\Gamma_\H$ is the amplitude of the radiation spectrum at
the hydrogen Lyman edge. Since $\bar\tau$ scales approximately as
$\tau\propto (\Omega_B h^2)^2/h \Gamma_\H$ (Rauch \etal 1997), we would
obtain the same results by keeping the original value of $\Gamma_\H$
from Haardt \& Madau but increase $\Omega_B h^2$ from 0.0125 to 0.0177,
which is still well within the range allowed by nucleo-synthesis.

At several output times we compute simulated spectra along lines of
sight through the simulation box. Each spectrum is convolved with a
Gaussian with FWHM = 8 km s$^{-1}$, then re-sampled onto pixels of
width 3 km s$^{-1}$ to mimic the instrumental profile and
characteristics of the HIRES spectrograph on the Keck
telescope. Artificial noise is introduced by adding a Gaussian random
signal with zero mean, and standard deviation $\sigma=0.02$ to every
pixel (\ie a SNR of 50 for pixels at the continuum). The absorption
lines in these mock observations are then fitted with Voigt profiles
using an automated version of VPFIT (Carswell \etal 1987).

\label{sect:simulation}
\begin{figure}
\setlength{\unitlength}{1cm}
\centering
\begin{picture}(8,13)
\put(-7., -9.5){\includegraphics{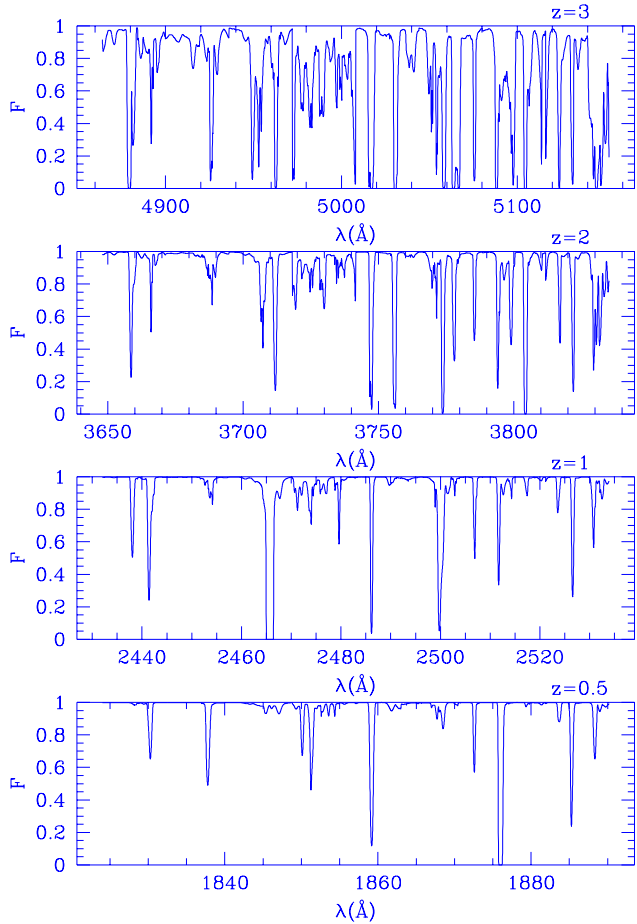}}
\end{picture}
\caption{Simulated spectra at redshifts 3,2,1 \& 0.5 (top to
bottom). We have combined eight spectra through the simulation box to
extend the wavelength coverage. The chosen sight-lines are the same at
each redshift. The wavelength range is chosen so as to show the same
comoving region at every redshift. }
\label{fig:spectra}
\end{figure}
\begin{figure}
\resizebox{\columnwidth}{!}{\includegraphics{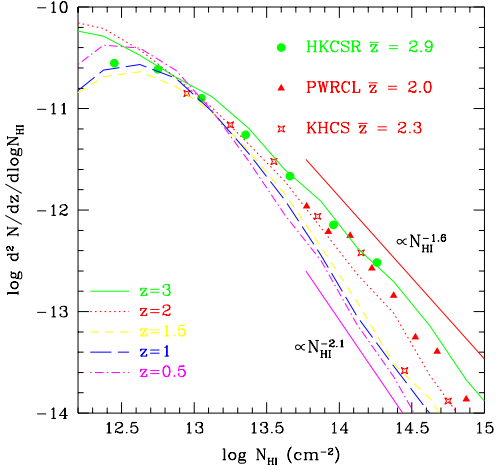}}
\caption{Evolution of the column density distribution with redshift,
for $z=3$, 2, 1, 1.5 and 0.5. Distributions for $z=2$ and $z=3$ are
from the higher resolution ($L=5.5$~Mpc) simulation, the others are for
the lower resolution ($L=22.22$~Mpc) simulation. Observational data
from Hu \etal (1995, HKCSR), Petitjean \etal (1993, PWRCL) and Kim
\etal (1997, KHCS) are shown for comparison, and their mean redshifts
are indicated. Scalings $d^2N/dz/dN_\H\propto N_\H^{-1.6}$ and
$N_\H^{-2.1}$ are shown.}
\label{fig:fig1}
\end{figure}

\section{Results and discussion}
\begin{figure*}
\setlength{\unitlength}{1cm}
\centering
\begin{picture}(17,12)
\put(1, -4){\includegraphics{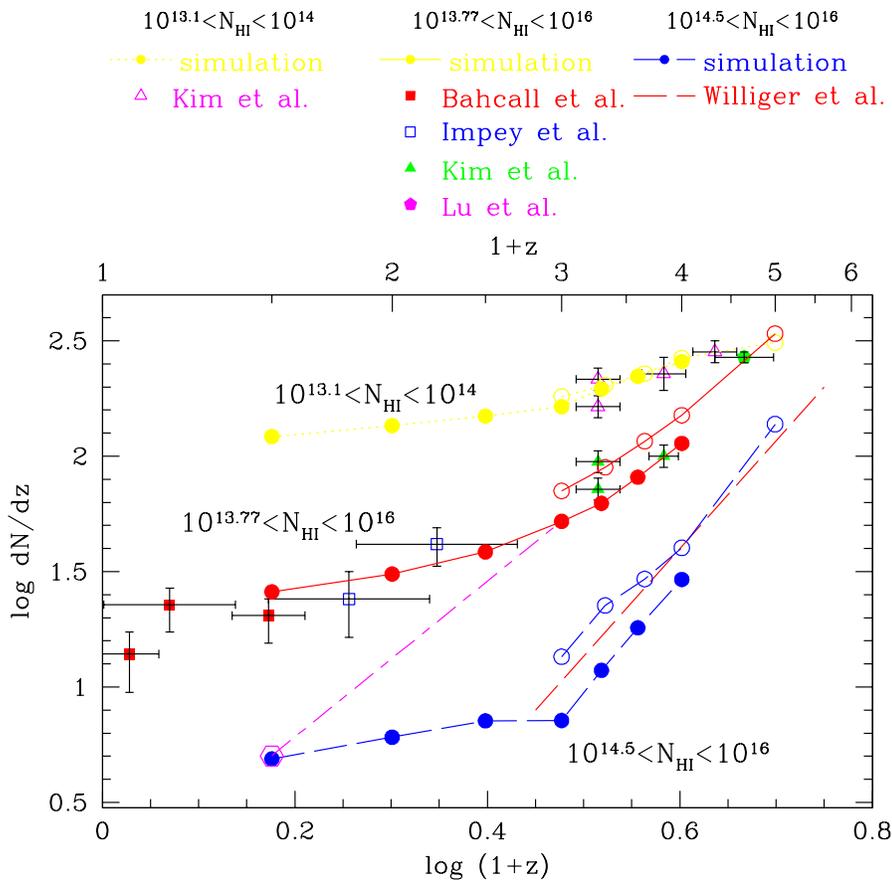}}
\end{picture}
\caption{Evolution of the number of lines with given range in column
density with redshift from simulations compared against observed
evolution. Column density cut $10^{13.1}$cm$^{-2}\le N_\H\le 10^{14}$
cm$^{-2}$-- simulations: circles connected with dotted line; data: open
triangles (Kim \etal 1997). Column density cut $10^{13.77}$ cm$^{-2}\le
N_\H\le 10^{16}$ cm$^{-2}$-- simulations: circles connected with solid
line; data: filled squares (Bahcall \etal 1993), open squares (Impey
\etal 1996), filled triangles (Kim \etal 1997), filled pentagon (Lu
\etal 1996). Column density cut $10^{14.5}$ cm$^{-2}\le N_\H\le
10^{16}$ cm$^{-2}$ -- simulation: circles connected with long dashed
line; data: long dashed line shows evolution from Williger \etal
(1994). Large filled circles are simulation results for the low
resolution, large box size, run, open circles are for a higher
resolution, smaller box size, run. Large open pentagon: re-analysis of
simulation at $z=0.5$, but imposing the ionizing background appropriate
to $z=2$. Data were taken from figure~2 in Kim \etal (1997), except for
the Wiliger \etal (1994) data. }
\label{fig:fig2}
\end{figure*}
We show in figure~\ref{fig:spectra} examples of simulated spectra at
$z=3$, 2, 1 \& 0.5 for the $L=22.22$ Mpc lower resolution
simulation. Fluctuations in neutral hydrogen density, caused by gas
tracing dark matter potential wells, produce absorption features
similar to those in observed spectra. At low redshifts, there are large
regions of the spectrum with very low absorption. These regions are
separated by prominent absorption features, most of which are just a
single strong line. At higher redshifts, there is considerable
absorption over most of the spectrum and many lines are blended. Many
of the strong lines at $z=0.5$ can be traced back to higher
redshifts. There is a clear decrease in the comoving number of lines
with decreasing redshift.

The evolution of the column density distribution with redshift is
illustrated in figure~\ref{fig:fig1}. There is clear evolution in the
simulated column density distribution with redshift. The rate of change
depends on column density, with higher column density lines undergoing
stronger evolution leading to steepening of the distribution. At $z=2$,
the column density distribution is $\propto N_\H^{-1.6}$ whereas this
has steepened to $\propto N_\H^{-2.1}$ at $z=0.5$ (see
figure~\ref{fig:fig1}). The rate of evolution also depends on redshift,
with considerably stronger evolution at higher redshifts. The number of
weak \lya lines ($\le 10^{13.1}$ cm$^{-2}$) remains approximately
constant. At redshifts 3 and 2, there is good agreement between the
simulated column density distribution for our higher resolution
simulation ($L=5.5$~Mpc) and the observed one.

The drop in the number of lines with redshift can be quantified by
counting the number of lines within a given column density range. The
simulation results are compared to observations in
figure~\ref{fig:fig2}. The simulations reproduce well the number of
lines at a given redshift for all three column density cuts.
Crucially, they also match very well the observed number of lines at
low redshift. Consequently, the hierarchical picture of galaxy
formation in a critical density universe, coupled with the observed
evolution in the quasar luminosity function, can explain the observed
evolution of the number of \lya lines over the entire observed redshift
range.

We have re-analysed the $z=0.5$ output time after increasing the
imposed ionization flux from its $z=0.5$ value to the value appropriate
to $z=2$. The number of lines with $10^{13.77}$ cm$^{-2}\le N_\H\le
10^{16}$ cm$^{-2}$ is shown by the open pentagon in
figure~\ref{fig:fig2}. This point falls onto the extrapolation for the
number density evolution for $z\ge 2$. Consequently, the dominant
reason for the higher number of lines at low $z$ compared to what would
be expected by extrapolation from high $z$, is the decrease in ionizing
flux from $z=2$ to $z=0$, itself a consequence of the evolution of the
quasar luminosity function.

An estimate of the reliability of these simulations can be obtained by
comparing the two simulations run at different resolutions. The higher
resolution simulation produces more lines at all column densities, but
the difference between the two simulations is well within the error
bars of the observational results. This gives us confidence that we can
reliable predict the number of lines from these simulations.

In summary: our numerical simulations show that the properties of the
\lya forest are in excellent agreement with what is expected in a cold
dark matter universe with a photo-ionizing background dominated by
quasar light. In particular, our simulations show that the observed
decrease in the rate of evolution of \lya absorption lines at $z\le 2$
can be explained by the steep decline in the photo-ionizing background
resulting from the rapid decline in quasar numbers at low redshift.

\section*{Acknowledgements}
TT acknowledges partial financial support from an EC grant under
contract CT941463 at Oxford University. APBL thanks PPARC for the award
of a research studentship and GPE thanks PPARC for the award of a
senior fellowship. We thank Martin Haehnelt for stimulating
discussions.

{}
\end{document}